\documentclass[preprint]{aastex}
\newcommand{\etal}{{et al.}}
\newcommand{\eg}{{\it e.g.,}}
\newcommand{\ie}{{\it i.e.,}}
\begin{document}

\title{Stellar Populations in the Host Galaxies of Mrk\,1014, IRAS\,07598+6508, and Mrk\,231\footnotemark[1]}

\footnotetext[1]{Based on observations made with the NASA/ESA Hubble
Space Telescope, obtained from the data archive at the Space Telescope
Science Institute, which is operated by the Association of Universities
for Research in Astronomy, Inc., under NASA contract NAS 5-26555.}

\author {Gabriela Canalizo\altaffilmark{2} and Alan Stockton\altaffilmark{2,3}}
\affil{Institute for Astronomy, University of Hawaii, 2680 Woodlawn
 Drive, Honolulu, HI 96822}

\altaffiltext{2}{Visiting Astronomer, W.M. Keck Observatory, jointly operated
by the California Institute of Technology and the University of California.}

\altaffiltext{3}{Visiting Fellow, Research School of Astronomy and Astrophysics, Australian National University.}

\begin{abstract}
We present deep spectroscopic and imaging data of the host galaxies of
Mrk\,1014, IRAS\,07598+6508, and Mrk\,231.   These objects form part of
both the QSO and the ultraluminous infrared galaxy (ULIG) families, and may 
represent a transition stage in an evolutionary scenario.   Our imaging
shows that all three objects have highly perturbed hosts with tidal tails 
and destroyed disks, and appear to be in the final stages of major mergers.
The host galaxies of the three objects have spectra typical of E+A galaxies,
showing simultaneously features from an old and a young stellar component. 
We model spectra from different regions of the host galaxies using \citet{bru96}
spectral synthesis models using two component models including
an old underlying population and recent superposed starbursts.   Mrk\,1014 
has intense star formation concentrated in a large knot $<2$ kpc from the
nucleus, along the leading edge of the tidal tail, and in several knots 
scattered around the host.   The starburst ages in these regions range from
180 to 290 Myr.  IRAS 07598+6508 has multiple knots of star formation 
concentrated in two regions within 16 kpc of the QSO nucleus, with ages
ranging from 30 to 70 Myr; the host galaxy shows an older population in other
regions.   Mrk\,231 shows a wider range of starburst ages, ranging from
42 Myr in the arc 3 kpc south of the nucleus, to $\sim300$ Myr spread on 
a ``plateau'' $\sim 20$ kpc across around the nucleus, as well as a UV bright
region 12 kpc south of the nucleus, which is apparently a region of currently
active star formation.
Our results indicate a strong connection between interactions and vigorous 
bursts of star formation in these objects.   We propose that the 
starburst ages found are indicative of young ages for the QSO activity.
The young starburst ages found are also consistent with the intermediate 
position of these objects in the far infrared color-color diagram. 
\end{abstract}

\keywords{galaxies: interactions---galaxies: infrared---galaxies: 
starburst---galaxies: evolution---quasars: individual 
(Mrk\,1014, IRAS\,07598+6508, Mrk\,231)}

\section{Introduction}

There has long been considerable circumstantial evidence that at
least some luminous, low-redshift QSOs are the result of strong interactions 
or mergers of galaxies 
\citetext{\citealp[\eg][]{too72,gun79,sto82,hut92,mcl94}; see \citealp{sto99}
for a review}.
However, a concrete suggestion for an evolutionary scenario for such 
objects was lacking until \citet{san88a} showed
that ultraluminous infrared galaxies (ULIGs), virtually all of which
are compelling examples of ongoing mergers, had bolometric luminosities 
and space densities similar to those of QSOs.   These similarities 
suggested the possibility that ULIGs are dust-enshrouded QSOs which, after 
blowing away the dust, become classical QSOs.   

\begin{figure}[h]
\epsscale{0.6}
\plotone{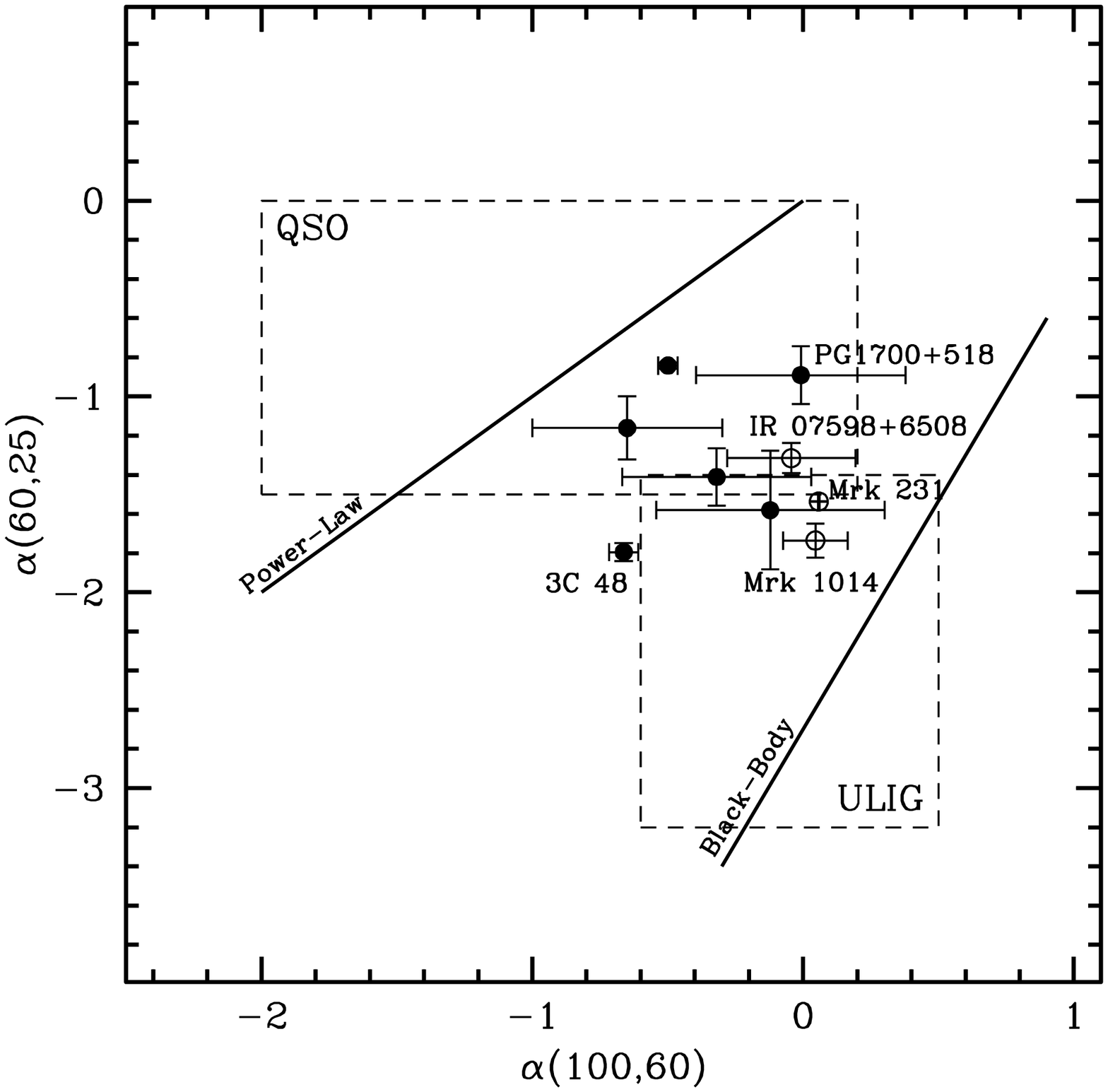}
\figcaption{Far infrared spectral index plot (adapted from L\'{\i}pari 1994).
Different kinds of objects such as QSO/Seyferts and
ultraluminous IR galaxies, occupy fairly well defined regions in the diagram
(see, \eg\ Neugebauer \etal\ 1985).   The objects in our sample 
have an intermediate position between these populations.   The open circles
show the positions of the three objects that are the subject of this paper.
\label{firplot}}
\end{figure}

If this hypothesis is correct, one should be able to observe examples of 
objects that are at intermediate stages of this evolutionary sequence.
We are conducting a study of a sample of low-redshift objects that may be 
in such a transitionary state.  These objects are recognized as bona-fide QSOs
and are found at an intermediate position in a far 
infrared (FIR) color-color diagram between the regions occupied by 
typical QSOs and ULIGs (see Fig.~\ref{firplot}).    
FIR color--color diagrams have been used as tools to
detect and discriminate different types of activity in the nuclear and
circumnuclear regions of galaxies.   Different kinds of objects such as
QSO/Seyfert, starbursts, and powerful IR galaxies, occupy fairly well defined 
regions in the diagram \citep[see, \eg][]{neu85,deG87,tan88,lip94}.
With deep imaging and spectroscopic observations of the host galaxies, 
we are attempting to construct interaction histories for each of these 
``transition'' objects.

If strong interactions triggered the QSO activity and induced starbursts, one 
might expect both events to occur roughly simultaneously, since both are 
plausibly dependent on gas flows to the inner regions.
Thus, we are placing these objects on an age sequence by
measuring the time elapsed since the last major starburst event.
This age sequence along with interaction histories
can help us answer the question of whether the intermediate position
of these objects is indicative of evolution from the ULIG to the classical
QSO population, or whether it
simply indicates a range of characteristics in QSOs.

Our sample is drawn from the \citet{neu86}, \citet{low88}, and \citet{cle96}
samples of {\it Infrared Astronomical Satellite\footnotemark[2]} ({\it IRAS}) 
objects, and
it consists of those objects which have: (1) a luminosity above the cutoff
defined for quasars by \citet{sch83},
\ie\ $M_{\rm B} = -23$ for 
$H_{0}=50$ km\,s$^{-1}$ Mpc$^{-1}$ (or $M_{\rm B} = -22.1$ for 
$H_{0}=75$ km\,s$^{-1}$ Mpc$^{-1}$),
(2) a redshift $z\leq0.4$,
(3) a declination $\delta\geq-30\arcdeg$,
(4) firm {\it IRAS} detections at $25 \mu$m, $60 \mu$m, and $100 \mu$m, and
(5) a position in the FIR color--color diagram which is intermediate between
  the ULIG and QSO loci (Fig.~\ref{firplot}).  
Although Mrk\,231 just misses the luminosity threshold given above, its
active nucleus is 
known to suffer heavy extinction (see \S\ref{mrk231}), apart from which
it would clearly be a member of the sample. We know of no other objects
satisfying the other criteria for which this is true.  We have therefore
chosen to include it for the present, although it may be approriate to exclude
it from some of the analyses of the whole sample, which will be presented
in a subsequent paper.

\footnotetext[1]{The Infrared Astronomical Satellite was developed and 
operated by the US National Aeronautics and Space Administration (NASA),
the Netherlands Agency for Aerospace Programs (NIVR), and the UK Science
and Engineering Research Council (SERC).}

So far, we have presented results for two of the nine objects in the sample:
3C\,48 \citep[ hereafter CS2000]{can00},
an ongoing merger near the peak of starburst activity; 
and PG\,1700+518 \citetext{\citealp{can97,sto98}, hereafter CS97 and SCC98; 
see also \citealp{hin99}},
where a tidally disturbed companion with a dominant 85 Myr old post-starburst 
population may be in the process of merging with the host galaxy.
In this paper we present the results for three additional objects:
Mrk\,1014, IRAS\,07598+651, and Mrk\,231. We assume $H_{0}=75$ km s$^{-1}$ 
Mpc$^{-1}$ and $q_{0}=0.5$ throughout this paper, so that the projected
physical length subtended by 1\arcsec\ is 2.43 kpc for Mrk\,1014, 2.26 kpc for
IRAS\,07598+651, and 0.77 kpc for Mrk\,231.

\section{Observations and Data Reduction \label{mrkobs}}

Spectroscopic observations for the three objects were carried out using the 
Low-Resolution Imaging Spectrometer \citep[LRIS;][]{oke95}
on the Keck II telescope.
For IRAS\,07598+6508, we used a 600 groove mm$^{-1}$ grating blazed at 
5000\,\AA\ yielding a dispersion of 1.28\,\AA\ pixel$^{-1}$.   For Mrk\,231 
and Mrk\,1014, we used a 300 groove mm$^{-1}$ grating blazed at 5000\,\AA\ 
with a dispersion of 2.44\,\AA\ pixel$^{-1}$.  The slit was 1\arcsec\ wide, 
projecting to $\sim$5 pixels on the Tektronix 2048$\times$2048 CCD.
We obtained two or three exposures for each slit 
position, dithering along the slit between exposures.
Table \ref{journal} shows a complete journal of observations, with
specification of the slit positions, and total integration times.

The spectra were reduced with IRAF, using standard reduction procedures.
After subtracting bias, dividing by a normalized halogen lamp flat-field
frame and removing sky lines, we rectified the two-dimensional spectra and 
placed them on a wavelength scale using the least-mean-squares fit of cubic 
spline segments to identified lines in a Hg-Kr-Ne lamp.
We calibrated the spectra using spectrophotometric standards from \citet{mas88}
observed with the slit at the parallactic angle.  The distortions in the 
spatial coordinate were removed with the IRAF {\it apextract} routines.
For each slit position, we had two or three individual frames;  we 
averaged the spatially corrected spectra using the IRAF task {\it scombine}.
We then corrected the spectra for Galactic extinction, using the values 
given by \citet{sch98}.

Since we were aiming to observe the youngest populations in the host
galaxies of these objects, we chose the slit positions based on previously 
obtained color maps of the host galaxies.  
We obtained imaging data for the three objects with the University of 
Hawaii 2.2~m telescope as specified in Table \ref{imaging}.
Observations with the $U'$ filter (centered at 3410 \AA\ with a bandpass 
of 320 \AA ) sample
the spectral energy distribution (SED) of galaxies on the short side of 
both the 4000 \AA\ break and the Balmer limit.   This region is very 
sensitive to the age of the stellar population, and will be brightest 
in regions of very recent star formation (ages $\lesssim 100$ Myr), 
as well as regions with scattered QSO
continuum.   Observations in $B-$band sample 
the region redwards of the Balmer limit, where $\sim$A-type stellar
populations are expected to peak, while longer wavelength optical 
and near infrared images will map the distribution of late-type
stellar populations.    The $U'\!-\!B$ color maps will then
highlight the regions of most recent star formation while $B\!-\!R'$
will point to the somewhat older ($\gtrsim 200$ Myr) populations.  Thus 
our slit positions generally cover the regions brightest in these color
maps.
In general, the spectra from each slit position were subdivided into regions
corresponding to some of the main features observed in the optical 
ground-based images, and one-dimensional spectra were extracted by summing
pixels corresponding to the regions of interest. 


The spectra of regions close to the QSO nucleus were contaminated by 
scattered QSO light.   The scattered light was removed by subtracting 
from each region a version of the quasar nuclear spectrum,
scaled to match the broad-line flux.  
In the case of Mrk\,1014, none of our slit positions went through
the QSO, so we obtained a separate 200 s exposure of the nucleus. 

Spectra from those slit positions which actually went through the QSO nucleus 
suffered from strong light scattering within the spectrograph, particularly
in the spectral region around H$\alpha$.
Therefore we were unable to obtain spectra of regions closer than 
$\sim2$\arcsec\ from the QSO nucleus for these slit positions.

{\it HST} WFPC2 images of the three objects were obtained from the {\it HST} 
data archive.   We used three 600 s WFC2 images of Mrk\,1014 in the F675W 
filter, one 400 s and two 600 s PC1 images of IRAS\,07598+6508 in the F702W 
filter, and two 1100 s PC1 images of Mrk\,231 in the F439W filter.
Most cosmic rays were removed by subtracting a median image from each of
the individual frames, then thresholding the difference at a
3$\sigma$ level, setting points above this threshold to the median of
the difference image.  Pixels near the position of the peak of the QSO
were excluded from this process.  The corrected difference image was then
added back to the median image, giving a corrected version of the original
image.   The few
cosmic rays within the relevant region that escaped this process were 
removed manually with the IRAF task {\it imedit}.  
In the case of Mrk\,231, where only two images were available, we first 
subtracted one from the other, and proceeded as above.
The procedure was repeated, interchanging the images.
All the corrected images for each object were then averaged.

\subsection{Modeling Spectra \label{modeling}}

We use \citet{bru96} isochrone synthesis models 
to fit the spectra of the host galaxies.   We will see in the following
sections that the three objects which are the subject of this paper show strong
evidence of having undergone some major tidal interaction.
As we have described in our previous work (CS97, CS2000),
spectra of the host galaxies of such objects show features from both young
(\eg\ strong Balmer lines) and old (\eg\ \ion{Mg}{1} {\it b} absorption)
stellar populations, and can usually be fitted satisfactorily
by a two-component model.  This model includes an old underlying stellar 
population,  presumably the stellar component present prior to interaction, 
and a younger instantaneous burst model, presumably produced as a result of 
the interaction.   A population with no age dispersion will be a reasonable
approximation of the actual starburst as long as the period during which the
star formation rate was greatly enhanced is short compared to the age of 
the population itself.

We have also noted previously (SCC98) that the age of the superposed
starburst is remarkably robust with respect to the different assumptions 
about the nature of the older stellar component.
Thus, we select a reasonable old underlying population (with certain
variations as described in each section below) and assume that the 
same underlying population is present everywhere in the host galaxy.
To this population we add instantaneous burst, \citet{sca86}
initial mass function, solar metallicity models of various ages.  
We then perform a $\chi^2$ fit to the data to determine the scaling of 
each component and the age of the most recent starburst.  The errors in
the starburst ages that we quote are estimated by noting the youngest
and oldest best fits for which $\chi^{2}$ changes by 15\% with respect
to the minimum value (see CS2000 for details).

In some cases, stellar absorption features (most often the Balmer lines) 
were contaminated
by emission coming from the extended narrow emission line region around
the QSO.   In some cases, we subtracted a scaled synthetic spectrum of
the recombination lines assuming Case~B.   However, in 
calculating $\chi^{2}$ for the model fitting, we generally excluded
those lines that suffered most from contamination.  All spectra are displayed
as observed (\ie\ without line subtraction), unless otherwise specified.

As we discussed in CS2000, because of our limited spatial resolution 
(generally $1\arcsec \times 1\arcsec$) and projection along the line of 
sight, we are likely observing the integrated
spectrum of several starbursts of different ages, and the age we determine
will be somewhat older than the youngest starbursts.   Therefore
the ages we report should be regarded as upper limits to the most recent
episodes of star formation along the line of sight. 

In objects with recent starbursts, the effect of reddening by dust is an
obvious concern.  However, studies of low-redshift AGNs and ULIGs at 
millimeter and submillimeter wavelengths indicate that dust is generally
heavily concentrated within $\sim1$ kpc of the nucleus \citep{and99, bry96}.
In addition, have found (SCC98) that even in a case where the
optical---IR spectral index is strongly affected by dust, the stellar ages 
from spectral features in the rest-frame 3200--5200 \AA\ region remain
fairly robust.  This relative insensitivity to dust can be attributed to
the fact that we are largely dealing with dust that is intermixed with the
stars and that we preferentially observe regions with low extinction and, thus,
low reddening.  Even in regions where dust along the line of sight is 
significant, the reddening in our observed bandpass is likely to be largely 
compensated by blue light scattered into our line of sight.

\section{Mrk\,1014 \label{mrk1014}}

Mrk\,1014 ($z=0.163$) is a luminous (M$_{B} = - 23.9$), infrared loud 
\citep[\eg][]{san88b}
radio-quiet QSO which shows a luminous host galaxy.     
The host galaxy has two large ``spiral-like arms'' (\citealt{mac84};
see Fig.~\ref{mrk1014mos} and Fig.~\ref{mrk1014hst}).
Its spectrum indicates a mixture of old and young stars 
\citep{mac84,hec84,hut90}.

Recently, \citet{nol00},
in a spectroscopic survey of 26 RLQs, RQQs, and
radio galaxies, observed the host galaxy of Mrk\,1014 with the Mayall 4 m
telescope at Kitt Peak National Observatory and with the 4.2 m William Herschel
telescope at La Palma.   They modeled the spectra and determined an age of 
12 Gyr for the host galaxy of Mrk\,1014.
Their approach is exactly opposite to ours (see \S\ref{modeling}): 
they fix the age of a possible young single starburst population to 0.1 Gyr 
and let that of a second, older single
starburst population (what we would call the ``underlying population'') vary. 
Their reason for including the 0.1 Gyr population is to account for
``the spectral shape of the blue light'' which they attribute to either
a recent burst of star formation or contamination of the slit by scattered 
light from the QSO nucleus.   Their results will be discussed further in
\S\ref{smrk1014spec}.

\subsection{Stellar Populations \label{smrk1014spec}}

The host galaxy of Mrk\,1014 shows stellar absorption features with 
redshifts remarkably close to those of the QSO broad and narrow emission
lines ($z_{\rm QSO}=0.1634$, as measured from our spectrum).  

We have obtained and modeled spectra of different regions in the host
galaxy of Mrk\,1014, and we shall refer to them according to their label
in Fig.~\ref{mrk1014slit}.
We have chosen a 10 Gyr old population with an exponentially declining star 
formation rate with an e-folding time of 3 Gyr as an old underlying 
population.   This model fits reasonably well the spectrum of a galaxy 
60\arcsec\ west-southwest of the QSO at the same redshift and only slightly
smaller than the ``bulge'' (see below) of Mrk\,1014.
(This galaxy was in our slit while obtaining the 200 s exposure
of the QSO).   
The assumption that the host galaxy (or parent galaxies) of Mrk\,1014
had a similar star formation history to this galaxy need not be 
accurate as the precise model we use as the pre-existing population
makes little difference in the age determination of the starburst 
population (SCC98).   For comparison, we have tried using a generic 
elliptical galaxy spectrum, and a model with a longer e-folding time 
(\ie\ 5 Gyr) as underlying populations in the modeling of the spectra in
Mrk\,1014.  We obtain the same starburst ages, though slightly different 
flux contributions from the old population,  regardless of the model used.    

Figure~\ref{mrk1014mos} includes a $B'\!-\!R'$ color map of Mrk\,1014.
This image emphasizes those regions with a steeper blue continuum spectrum, 
peaking just redwards of the Balmer limit.   These regions are concentrated
mainly along the north edge of the tail (regions $a$ and $b$ in 
Fig.~\ref{mrk1014slit}), in a clump on the east end of the tail ($c$), 
directly east of the nucleus 
(not covered by our slits), and southwest of the QSO nucleus ($e$).

\begin{figure}
\epsscale{0.6}
\plotone{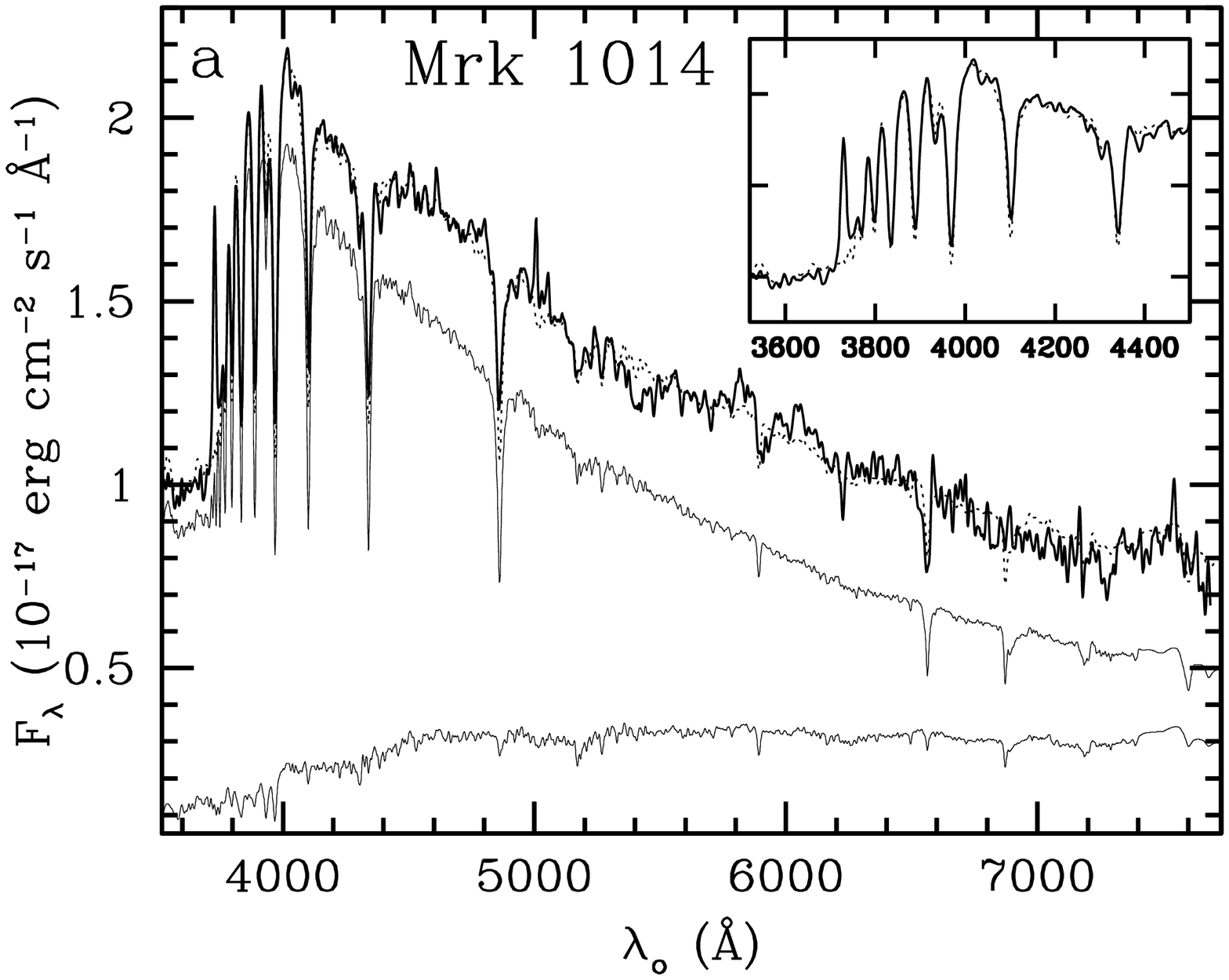}
\figcaption{Rest frame spectrum of region $a$ in the host galaxy of Mrk\,1014
(heavy line), underlying old stellar population (lower light line), 200 Myr
instantaneous starburst model (upper light line) and the $\chi^2$ fit of the 
sum of the two models to the data (dotted line).  The old model is a 10 Gyr 
old stellar population with an exponentially decreasing star formation rate 
with an e-folding time of 3 Gyr. The original data have been smoothed with 
Gaussian filters with $\sigma =2$ \AA.  The inset shows in detail the blue
part of the spectrum.  In this and the following figure, emission lines
come from the extended ionized gas around the QSO, and not from star forming
regions.  These lines have not been included in calculations of $\chi^2$.
\label{mrk1014young}}
\end{figure}

Spectroscopy of these regions confirms that they are indeed the youngest
stellar populations we find in the host, with ages ranging from
180 Myr in region $e$ to 290 Myr in region $d$.  Figure~\ref{mrk1014young} 
shows the spectrum of region $a$, with the best $\chi^{2}$ fit of the model
to the data superposed,  and the relative contributions of the 200 Myr 
starburst and the old underlying population.  The error in these ages is
typically $\pm 50$ Myr.   Figure~\ref{mrk1014hst}
shows compact knots at the positions of regions $a$ through $d$, and 
Planetary Camera {\it HST} images \citep{sur98}
show a larger and very bright blue knot at the position of region $e$.
Even though there are knots at the positions of $a$ and $b$, it is 
evident from colors and spectroscopy that there is recent star formation
all along the north edge of the tail (\ie\ between $a$ and $b$).

\begin{figure}
\epsscale{0.6}
\plotone{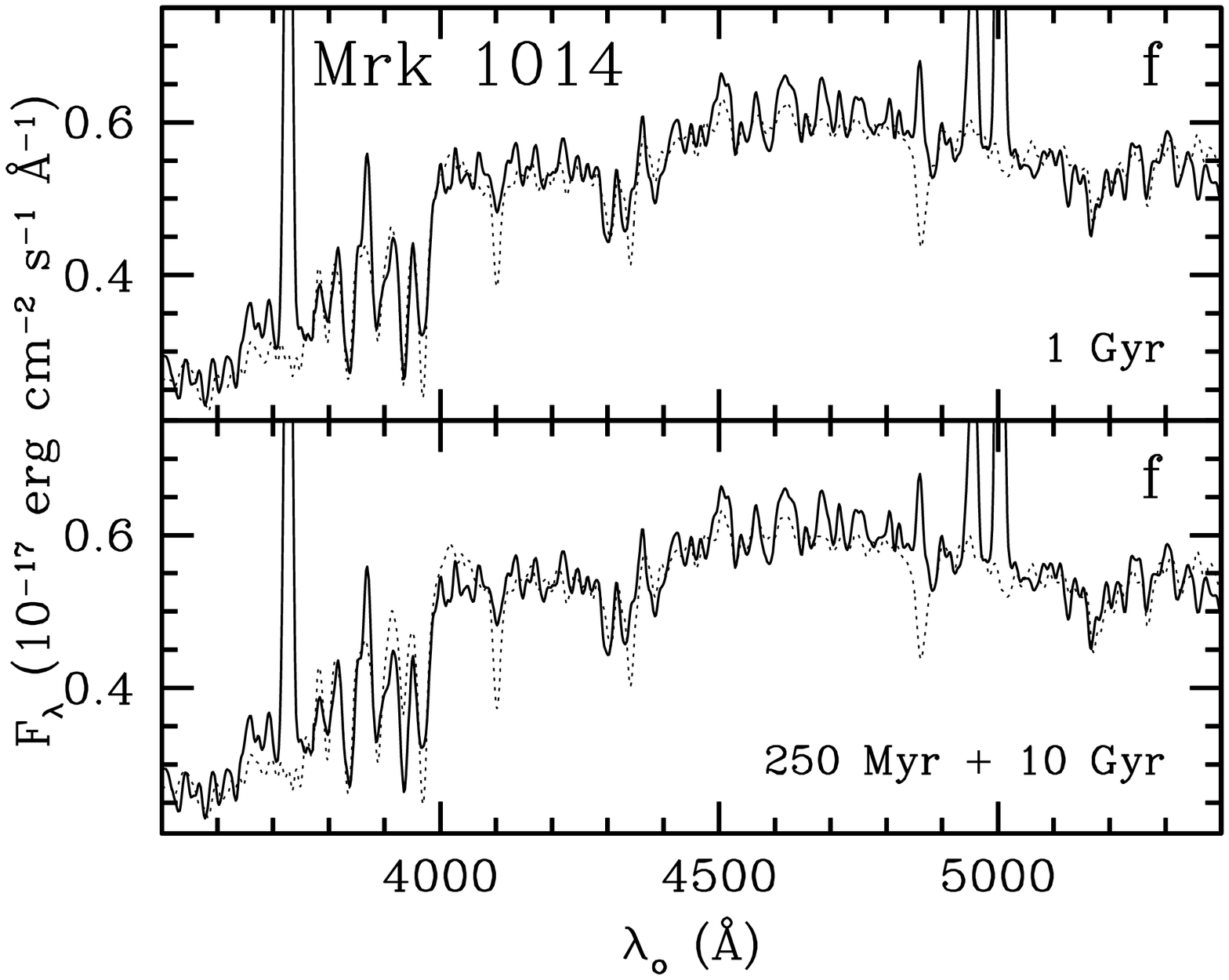}
\figcaption{Rest frame spectrum of region $f$ in the host galaxy of Mrk\,1014
(solid line in both panels).  The dotted line in the top panel is a 
single 1 Gyr instantaneous burst model.   The dotted line in the bottom 
panel is the best 
$\chi^2$ fit to the data of the sum of a 250 Myr instantaneous burst model
and an old population (see text for details).\label{mrk1014older}}
\end{figure}

Other regions of the host galaxy  appear redder in the $B'\!-\!R'$ color map
in Fig.~\ref{mrk1014mos}.   These regions, sampled by $f$ and $g$,
appear to be dominated by an older, $\sim1$ Gyr population, 
and to have very little,  if any, contribution from the old 
underlying population (Fig.~\ref{mrk1014older}, top panel).   
It is not entirely clear whether these are truly
intermediate age populations, or if they are simply, as in the case of 
region $h$, dominated by an old underlying population 
with a very small contribution from a younger population like those
found in regions $a$ through $e$.    We attempted to fit models with
the latter characteristics to the observed spectra.   The resulting fits
are reasonable, with minimum values of $\chi^{2}$ 10\% and 25\% larger than 
that obtained for a dominant intermediate-age population
for regions $f$ and $g$ respectively.    
The bottom panel of Fig.~\ref{mrk1014older} shows the spectrum of region $f$
with the best fit to the data of the sum of an old underlying
population and a 250 Myr instantaneous 
burst population.  While a reasonable fit, it does show significant 
discrepancies in the region around the 4000~\AA\ break and the 
\ion{Ca}{2} $K$ line.  

The potential presence of this intermediate age population suggests that a 
better fit for the younger populations might be achieved by adding a third 
component that accounts for this intermediate age component.   However the 
flux of the young starburst population in those regions is so dominant
(typically contributing 80\% of the total flux at rest-frame 5000 \AA),
that a third component makes a negligible difference to the fit.

Region $i$ shows very strong emission, which almost certainly comes from 
gas ionized by the
QSO rather than from star-forming regions.  The equivalent widths of the
emission lines here are greater than those of the emission lines elsewhere in
the galaxy by at least a factor of 5, and at least twice those of the 
QSO nucleus for [\ion{O}{3}].  The emission line ratios indicate a power-law
ionizing continuum \citep{vei87}.
This region is clearly seen as a large, discrete knot in Fig.~2$d$ of \citet{sto87}.
The gas has an approaching 
velocity of $180 \pm 50$ km s$^{-1}$  with respect to the stellar absorptions 
in that region.   The underlying spectrum is similar to that of region $h$.

Region $h$ also shows emission lines with approaching velocities of $\sim 200$
km s$^{-1}$ with respect to the stellar features, but this region shows
an additional, weaker emission component clearly visible in [\ion{O}{2}],
[\ion{O}{3}], and H$\beta$, blueshifted by $1150 \pm 50$ km s$^{-1}$ with 
respect to the stronger component.   The stellar population here seems to
be dominated by an old population with a small fraction of the flux coming
from younger stars.

We have added spectra from two different slit positions at region $j$
to improve the signal to noise in this very faint region.   This is part 
of the long extension on the west side clearly seen in the high contrast 
image in Fig.~\ref{mrk1014mos}.
We find a continuum with a red SED and a clear 4000 \AA\ break at a 
redshift close to, but slightly larger than that of the main 
body of the galaxy.   

Region $k$, which is very bright in the $R'$ image in 
Fig.~\ref{mrk1014mos}, would appear to be an extension of the east tail.   
However, its spectrum shows that it is a background galaxy at $z=0.5843$.   
Likewise, region $l$ on the west tail is a galaxy at $z=0.5857$.  
A third galaxy SE of the QSO ($\Delta\alpha = +11\farcs9$, $\Delta\delta =
-40$\arcsec) has a similar redshift as well ($z=0.5838$), so there seems 
to be a group or cluster of galaxies at this redshift.

The 200 s exposure spectrum of the elongated object 9\farcs2
west-southwest of the QSO shows narrow emission lines at the same redshift as 
the QSO superposed on a red continuum.   Fig.~2$d$ of \citet{sto87}
shows that there is strong emission just
north of this object coming from extended gas ionized by the QSO and not 
necessarily associated
with the object.  The spectrum is too noisy to distinguish stellar features,
so it is unclear whether this object is interacting with the system
or if it is a chance projection.   The near edge-on galaxy $\sim23$\arcsec\
west of the QSO, on the other hand, shows [\ion{O}{3}] emission confined to
the galaxy at $z=0.162$ in a
University of Hawaii 2.2 m telescope spectrum (Canalizo 2000, unpublished),
as suggested by \citet{sto87}
from their [\ion{O}{3}] imaging.

How do our results compare to those of \citet{nol00}?
As mentioned above, Nolan \etal\ use a fixed 0.1 Gyr starburst population.   
They determine that this population makes up 1.1\% of the
total luminous mass along the line of sight, and that the rest of the flux is
well characterized by a 12 Gyr instantaneous burst.   In contrast,
we find several regions of recent ($\sim0.2$ Gyr) star formation
where the star forming mass typically amounts to 12\%, and sometimes 
up to 30\%, of the total luminous mass along the line of sight.
One of our slit positions is very similar to one used by Nolan \etal\
(see Fig.~2 in \citealt{hug00}),
but our slit is narrower
and slightly closer to the nucleus.   As a way to compare, we added the
flux along the slit as Nolan \etal\ seem to have done, including the
background galaxy $k$, which is also in their slit.      
Even if we add up all the flux along this slit (subtracting the QSO
light, which amounted to 5\% of the total flux), we still find that the
star forming regions make up 8\% of the total luminous mass along the
line of sight.    Obviously, the difference in ages will lead to a 
smaller percentage for their choice of parameters.  So,  
we tried fixing our parameters to match theirs (\ie\ 12 Gyr +
0.1 Gyr populations), and this yields a 5\% by mass for the young
population, but the fit is much inferior to the ones discussed in
this section.   Their slit position may have fortuitously missed the 
major star forming regions, thus leading to this smaller percentage.
We have previously cautioned (CS2000) that different slit positions
can lead to different age determinations, and we emphasize the
importance of carefully selecting slit positions if one wishes (as we do)
to find the major starburst regions.
It is also far more difficult to obtain a reliable age for an older
population in the presence of a contaminating younger population than the
reverse.   

\subsection{Interaction History}

Images of Mrk\,1014 show a very prominent tail extending to the northeast,
reminiscent of the tidal tail of 3C\,48 (CS2000).   Like the tidal tail in
3C\,48, this tail has a number of small clumps (see {\it HST} image in 
Fig.~\ref{mrk1014hst}) 
of star formation, which are commonly found in merging systems.   The 
bulk of the bright portion of the tail appears to be dominated by an
intermediate age population ($\sim$1 Gyr) with a very small contribution, 
if any, from an old underlying population;
alternatively, it could be dominated by an older population, with some
flux scattered from the bright star-forming regions ($a$ through $d$)
or fainter small regions distributed along the host. 
However, the north edge of the tail
appears as a very sharp feature in the short wavelength images, and contains 
stellar populations
which are as young as those of the clumps.   As we find no redshift
variations along this edge, it would appear that we are observing the
tail nearly face-on, and that this is the leading edge where the material
has been compressed, thus producing star formation.   This, again, is
similar to the blue leading edge observed in 3C\,48 (Fig.~2$e$ in CS2000),
presumably sharper in Mrk\,1014 because of the lower inclination angle.
  
Both the ground-based and the {\it HST} images of Mrk\,1014 
(Figs.~\ref{mrk1014mos} and \ref{mrk1014hst}) show a long,
low-surface-brightness extension of the bright tail on the east of the
nucleus and arching towards the south (more evident after the removal of 
the background galaxy, $k$) 
as well a very extended faint secondary tail, rotationally symmetric to
the bright (primary) tail.   Each tail extends for as much as $\sim40$\arcsec\ 
or $\sim$100 kpc (note the inset in Fig.~\ref{mrk1014mos} showing the 
well-known local interacting system M\,51/NGC\,5195 at the same scale).  
Spectra of the secondary tail west of the nucleus indicates that the tail 
is made up by older stars, and this is consistent with this 
tail being visible in our $H$ (not shown) and $K'$ images.
No bright clumps of star formation are evident along the secondary tail;
this absence is not unusual as tidal dwarf formation appears not to be a 
ubiquitous process in mergers \citep{hib99}.
The {\it HST} image of the nucleus (see inset in Fig.~\ref{mrk1014hst})
shows a small extension on the south side which 
follows the direction of the secondary tidal tail.   A similar extension is
seen in {\it HST NICMOS} images by \citet{sco00}.
There appears to be a ``bulge''
or enhanced brightness area elongated roughly along the axis connecting 
the beginning of both tails (\ie\ NW--SE), with a half radius of 
$\sim4$ kpc.
Assuming a projected velocity of 300 km s$^{-1}$ on the plane of the
sky, the dynamical age for the tails is $\sim330$ Myr.
The tails are then older than every major post-starburst knot they contain,
consistent with the idea that the latter were formed after the tails
were first launched.   At the same time, the tails are dynamically younger
than the bulk of the stars that form them, though perhaps only slightly so 
if the stellar
populations observed in $f$ and $g$ are truly $\sim1$ Gyr old, instead of
being $\sim10$ Gyr with a small admixture of younger stars.

Mrk\,1014 shows some striking similarities to 3C\,48 (CS2000): the 
morphology of the (primary) tidal tail, the clumps of star formation along 
the tail as well as its blue leading edge, the relation of the starburst 
ages to the dynamical age of the 
tails (both ages $\sim200$ Myr younger in 3C\,48), and the clumpy extended
emission line region \citep{sto87}
with high velocity ($>1000$ km s$^{-1}$) components.   
As with 3C\,48, the data strongly suggest that 
Mrk\,1014 is the result of a merger of two galaxies of comparable size, both 
of which were disks in this case.
The starbursts in the main body of the host of Mrk\,1014, however, appear to 
be less intense and less widespread than those of 3C\,48.  

If the intermediate age (1 Gyr) population 
we see in regions $f$ and $g$ is real, it may be the relic of
a starburst ignited at an initial passage of the two interacting galaxies.
Indeed, we know of another system \citep[UN\,J1025$-$0040;][]{can00b}
where the interacting galaxies have a difference in starburst ages of this 
order,  possibly coincident with their orbital period.
If this were the case for Mrk\,1014, one might expect most of the star 
formation at the present to
be very strongly concentrated towards the nucleus, as most of 
the gas would have been driven towards the center starting $\sim1$ Gyr ago.

\section{IRAS\,07598+6508 \label{ir0759}}

The luminous, $z=0.148$, radio-quiet QSO IRAS\,07598+6508 was first detected 
by {\it IRAS}, identified as an AGN candidate by \citet{deG87},
and spectroscopically identified as a QSO by \citet{low88}.
The optical spectrum of the QSO is dominated
by extremely strong \ion{Fe}{2} emission \citep{law88,lip94},
and the UV spectrum shows low- and 
high-ionization broad absorption lines (BAL) extending to blueshifts of
5200 to 22000 km s${-1}$ \citep{hin95,bor92}.

Figures \ref{ir0759mos} and \ref{ir0759hst} show, respectively, ground-based 
and {\it HST} images of IRAS\,07598+6508 \citep[see][ for a PSF-subtracted
version of the {\it HST} image]{boy96}.
These images show two 
clumpy regions $\sim7$\arcsec\ south and southeast of the
QSO, bright in the PSF subtracted $U'$ image, but barely visible in $H$-band
images (not shown).   The great number of knots in these regions, presumably
OB associations, already argues for recent star formation.

\begin{figure}[!h]
\epsscale{0.6}
\plotone{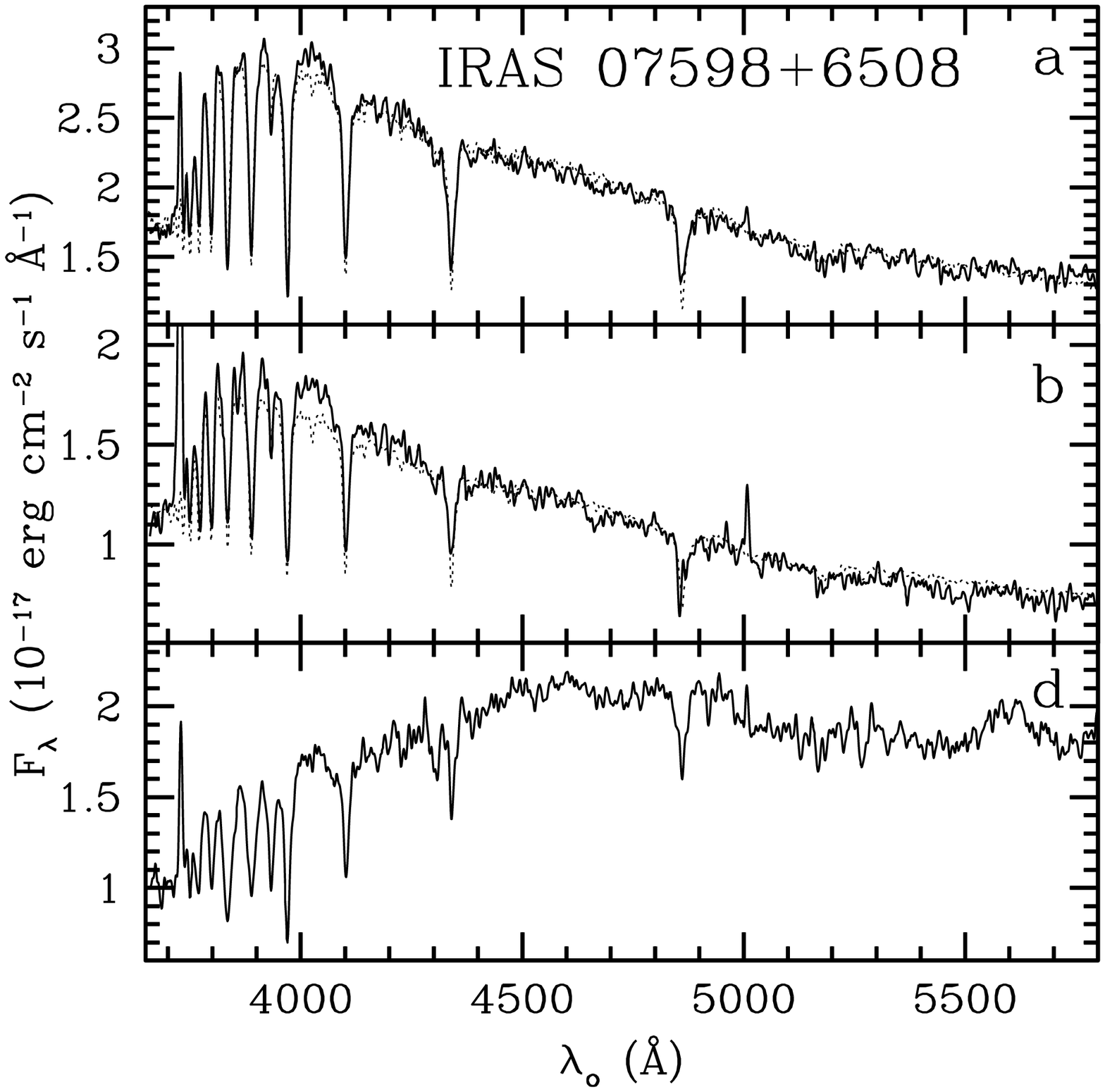}
\figcaption{Rest frame spectra of regions $a$, $b$, and $e$ (from top to 
bottom) in the host galaxy of IRAS\,07598+6508.  For panels $a$ and $b$, 
the solid lines are the observed spectra and the dotted lines are 
instantaneous burst models of 70 and 30 Myr, respectively.\label{ir0759spec}}
\end{figure}

Our slit positions cover these two regions, labeled $a$ and $c$ in
Fig.~\ref{ir0759slit}, as well as some of the fainter emission surrounding
the nucleus.  Spectra of these regions are shown in Fig.~\ref{ir0759spec}.
Regions $a$ and $c$ are fit by single starburst models of ages $70\pm15$ and 
$32\pm7$ Myr respectively.   The light from these starbursts dominates the
spectra as we do not find any significant contribution from an older 
component.   
Although single burst models fit the data better than two-component models,
there are still some discrepancies in the fit.   The observed spectra show
an excess with respect to the model in the region between 3900 \AA\ and
4100 \AA\ , apparently because the observed continuum is steeper in
this region.   To test whether the discrepancy could be an indication of
scattered QSO light, we subtracted QSO spectra with several different scalings
from the stellar spectra, but we were unable to obtain better fits.

In contrast to $a$ and $c$, the spectra of the regions closer to the 
nucleus (labeled $d$ and $e$ in Fig.~\ref{ir0759slit}) show an
older stellar population (Fig.~\ref{ir0759spec}, bottom panel).   
The SEDs of these
spectra are much redder and there is a clear, though not very prominent,
4000 \AA\ break.   We detect this
old population from $\sim2\arcsec$ to $\sim8\arcsec$ north of the
QSO and from $\sim2\arcsec$ to $\sim4\farcs 5$ south of the QSO.

We were unsuccessful in attempting to subtract the QSO scattered light closer
to the nucleus because of the strong light scattering along the slit
in these regions.   Therefore we are unable to determine whether the
populations closest to the nucleus are as old as those of $d$ and $e$
or if there might be a younger starburst concentrated around the nucleus.

Regions $a$ and $c$ have a slightly higher redshift than the
fainter emission around the QSO: $z_{a} = 0.1490 \pm 0.0001$ and 
$z_{c} = 0.1488\pm0.0001$, compared to $z_{d,e} = 0.1485 \pm 0.0002$ for 
the regions closer to the nucleus, but the difference is barely significant.  
Measuring a precise redshift for the QSO
is difficult because of the strong \ion{Fe}{2} emission contaminating the
broad emission lines, and the absence of narrow emission lines.
Values in the literature include $z_{\rm QSO} = 0.1488$ as measured from CO 
\citep{sol97}
and $z_{\rm QSO}=0.1483$ as measured from broad Balmer emission lines 
\citep{law88};
both are within 180 km s$^{-1}$ (and within the errors) of the values we 
measure for the host galaxy. 

IRAS\,07598+6508 has some limited extended narrow emission as seen in
the [\ion{O}{3}] image in Fig.~\ref{ir0759mos}.   The ionized gas seems
to be correlated with the stellar component:  the
velocity difference between emission and absorption lines in regions where 
both are present is never larger than 50 km $s^{-1}$ (notice the profile of
H$\beta$ in Fig.~\ref{ir0759spec}; emission is visible slightly redshifted
with respect to the absorption).   

Region $b$ shows the strongest emission
and appears brightest in the [\ion{O}{3}] image. The emission line spectrum
has an underlying blue continuum with stellar features.  
The 2-dimensional spectrum shows that this region is broken up into two
discrete clumps a little over 1\arcsec\ each, having some small
scale velocity structure.   The line flux ratios indicate that the south 
clump has lower ionization, and it may be an \ion{H}{2} region rather
than gas ionized by the QSO. A second,
fainter component with a velocity gradient from 0 to $\sim500$ 
km $s^{-1}$ apparently originating from the north clump extends 1\arcsec\ 
towards the south.

About 11\arcsec\ north of the QSO, we find another emission region 
(labeled $g$) which
is visible in the [\ion{O}{3}] image, but not in any of the broad band
images.   No stellar continuum is evident from the spectrum, either.
This region is also broken into two discrete, somewhat larger clumps, but 
in this case the clump closer to the nucleus has lower ionization.
Both clumps are at a lower redshift than those of region $b$, \ie\ 
$z_{\rm em} = 0.1483 \pm 0.0001$  {\it vs.}  $z_{\rm em} = 0.1491 \pm 0.0001$
in $b$.

Our deep $R'$ image (Fig.~\ref{ir0759mos}) shows a tidal tail extending 
from north to the east and arching towards the south
of the nucleus for $\sim22$\arcsec\ or $\sim50$ kpc.  The dynamical age
of this feature (assuming a projected velocity of 300 km $s^{-1}$) is 
$\sim160$ Myr, again older than the starbursts ages found for this object.
We have a very faint spectrum of this tail at region $f$.   The SED of
the spectrum is similar to that of $d$ and $e$, but the spectrum is too
noisy to show stellar features.    An [\ion{O}{2}] $\lambda 3727$ emission 
line is confined to this region (as seen in the 2-d spectrum) at a redshift 
of $z_{\rm em}=0.1487$, which is probably close to that of the stellar 
component as in other regions.

This tidal tail is strongly suggestive of a merger.
The fact that only one tail is evident may indicate that we are seeing
the result of a merger of a spiral with an elliptical galaxy.   This 
configuration, however, could also result from the merger of two
spiral galaxies, one of which is counter rotating to the relative orbit.
In mergers with such geometry, the gas and stars in the counter-rotating
disk are only slightly perturbed and are
not pulled into tidal bridges and tails \citep{too72,hib99}.
It is also possible that a second tail may not be evident because of projection
effects.   

Regions $a$ and $c$ may be part of the host galaxy with enhanced surface
brightness due to the recent star formation, or they may be remnants of
companion galaxies which have strongly interacted with the host galaxy.   
In either case it is clear that these regions are tidally disturbed and 
they will likely be 
completely mixed with the host within a few crossing times.

The two galaxies $\sim14$\arcsec\ south of the QSO are likely to be companion
galaxies since they are bright in our [\ion{O}{3}] image 
(Fig.~\ref{ir0759mos}),
but we do not have a spectrum to confirm this.   The colors for the SE galaxy
indicate that it may have recent star formation, if at the same redshift as
IRAS\,07598+6508.   We find two additional objects with the same 
redshift as IRAS\,07598+6508 that happened to fall in our slits: a faint 
emission line object $\sim20$\arcsec\ south-southwest of the QSO, and a 
bright galaxy with an absorption and emission line spectrum $\sim60$\arcsec\ 
southwest of the QSO.

\section{Mrk\,231 \label{mrk231}}

Mrk\,231 ($z=0.042$), often classified as a Seyfert 1 galaxy, is slightly
below the luminosity cutoff for QSOs defined by \citet{sch83}.
However, the nucleus is heavily reddened, with an estimated $A_V\approx2$
of foreground extinction 
\citetext{\citealp{bok77,lip94}; see also \citealp{goo94}}, so it would be
well above this threshold if it were unobscured.  The central plateau of
the host galaxy is off center with respect to the nucleus, and the presence
of tails to the north and south, as well as a low-surface-brightness
extension to the east clearly indicate a recent merger 
(Fig.~\ref{mrk231mos}).  Closer to the nucleus, {\it HST} imaging
shows a large number of stellar associations indicating recent star
formation (\citealt{sur98}; see Fig.~\ref{mrk231hst}).

As in the case of IRAS\,07598+651, the spectrum of 
the QSO is dominated by strong \ion{Fe}{2} emission \citep{lip94b}
and shows a peculiar low-ionization BAL system with velocities
up to $-7800$ km s$^{-1}$ \citep{smi95,fos95}.
Previous spectroscopy of the host galaxy indicates the presence of a 
young stellar population \citep{ham87}.

\begin{figure}[!h]
\epsscale{0.6}
\plotone{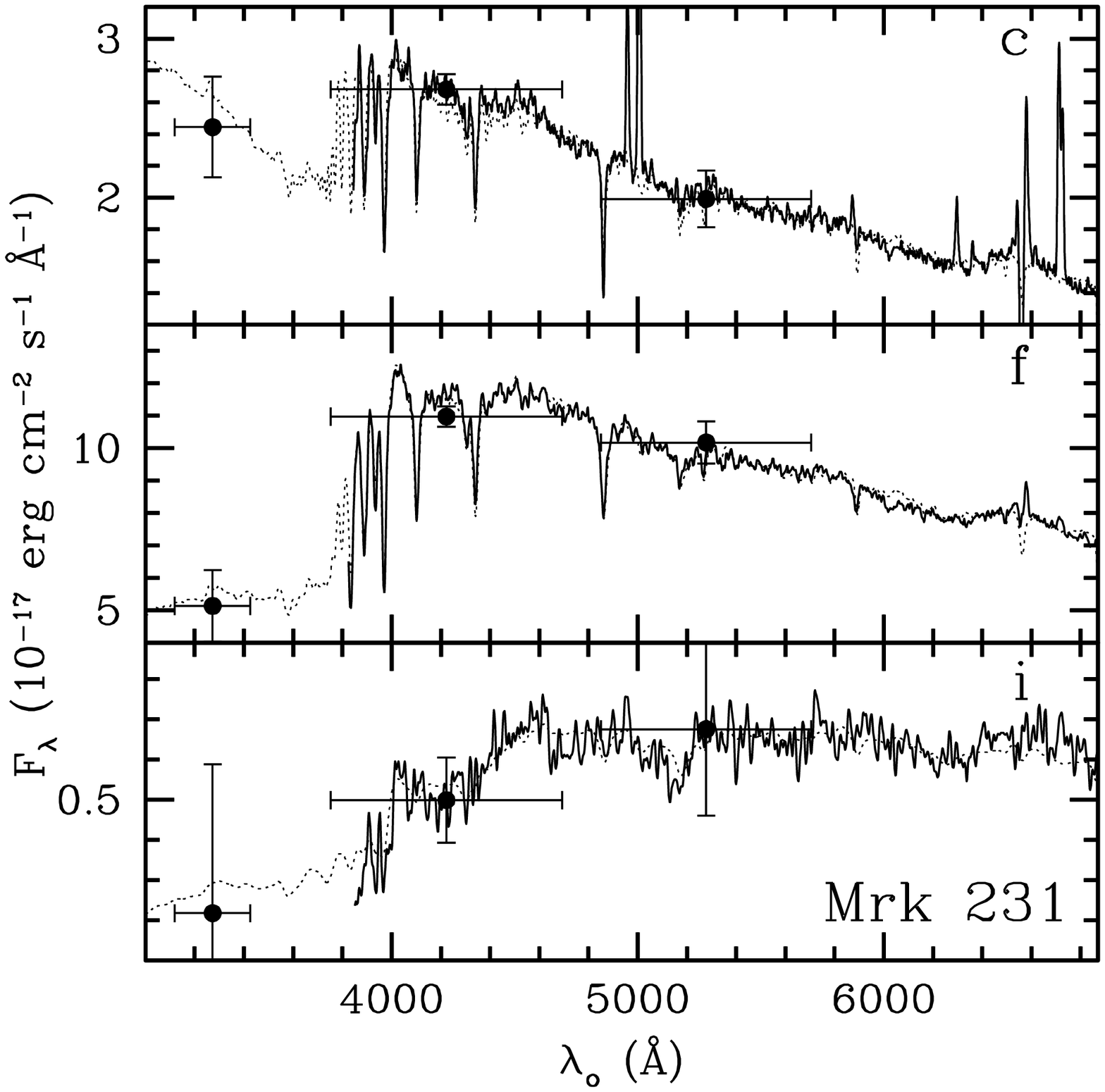}
\figcaption{Rest frame spectra of regions $c$, $f$, and $i$ (from top to 
bottom) in the host galaxy of Mrk\,231.   Each panel shows the observed
spectrum (solid line), the best $\chi^{2}$ fit of the model to the data
(dotted line), and the photometry of the region from our $U'$, $B$, and
$V$ images.   The error bars in the horizontal direction represent the 
de-redshifted bandpasses, while those in the vertical direction are 
1$\sigma$ random uncertainty in the flux.
Balmer emission lines of the spectrum in panel $c$ have been subtracted, 
assuming case B recombination.\label{mrk231spec}}
\end{figure}

We have obtained and modeled spectra of several regions in the host galaxy
as labeled in Fig.~\ref{mrk231slit}.  
In regions $k$ and $i$, we find no evidence for a significant young 
starburst population.  A 10 Gyr-old population with an exponentially 
decreasing 
SFR and an e-folding time of 5 Gyr fits this spectrum quite well, as
shown in the bottom panel of Fig.~\ref{mrk231spec}.   Therefore we use this
model as an underlying population to fit the rest of the spectra in the 
host galaxy.

Unfortunately, because of the poor sensitivity of LRIS shortwards of 4000 \AA ,
and the fact that Mrk\,231 has a lower redshift ($z = 0.042$) than the 
other two objects, we were unable to obtain the near-UV portion of 
the spectrum, which is the most helpful region in determining the ages of 
post-starburst populations.  We are left then with the region on the long 
side of 3800 \AA , where more than one combination of models can give
similar fits to the spectra.   

\begin{figure}[!h]
\epsscale{0.6}
\plotone{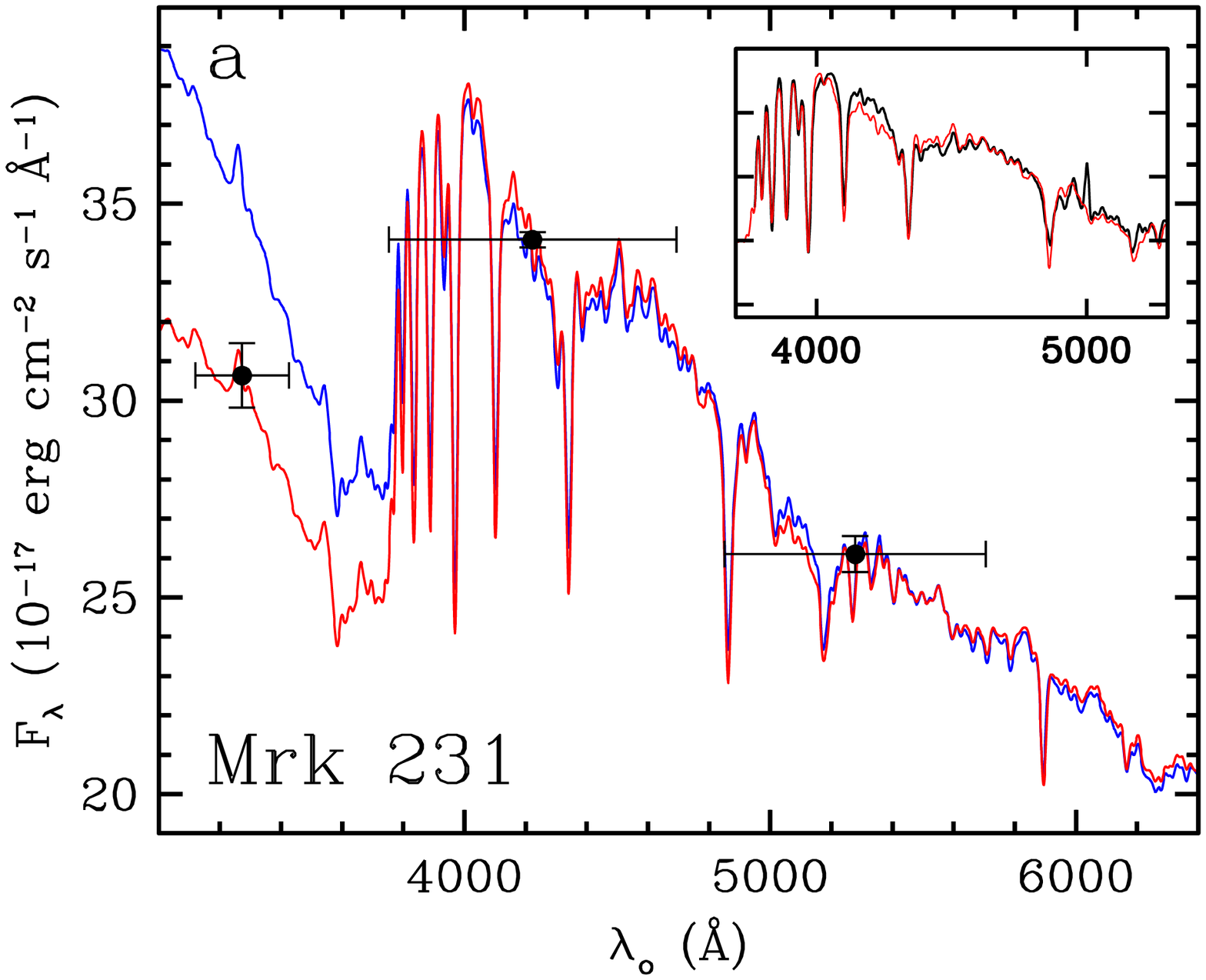}
\figcaption{Degeneracy in the modeling of spectra in the host galaxy of
Mrk\,231.   The main panel shows two spectral synthesis models giving very 
similar fits to the rest-frame spectrum of region $a$ redwards of 3800 \AA . 
The blue line is the sum of an old (10 Gyr with exponentially decaying
SFR) model and a 4 Myr instantaneous burst model that contributes 1\% of
the luminous mass.   The red line is the sum of the same old population and 
a 42 Myr instantaneous burst model that contributes 12\% of
the luminous mass.  There is only a 10\% difference in $\chi^{2}$ of the fit 
of each of these models to the data.   The inset shows the actual spectrum 
of region $a$ with (black line) and the older of the two models.
\label{degenerate}}
\end{figure}

Indeed, we found degenerate fits in some regions of the host galaxy,
particularly in those regions with the youngest populations and those where the
absorption lines were heavily contaminated by emission so that we could not
use Balmer lines to discriminate between models.
In such regions we obtained two $\chi^{2}$ minima, generally with
one corresponding to a model with a small contribution (\ie\ small percentage
of total mass along the line of sight) from a very young 
starburst, and the other to a model with 
a large contribution from a somewhat older starburst.   We illustrate the
problem in Fig.~\ref{degenerate}, where we have plotted two models resulting
from a 4 Myr and a 42 Myr old populations contributing, respectively, 1\%
and 12\% of the total luminous mass.
The models are nearly identical in
the spectral region redwards of 3800 \AA , and both are good fits to the data
(see inset in Fig.~\ref{degenerate}) with $\chi^{2}$ values for these models
differing only by 10\%. The models, however,  diverge quickly at shorter 
wavelengths.

In order to discriminate between the two ``best fits'' to the data, we
have measured photometry of the different regions ($a$ through $k$ as
indicated in Fig.~\ref{mrk231slit}) from our ground-based optical images.
We have been careful to measure fluxes only within the areas limited by
the slits, at similar resolutions, and subtracting scattered light from 
the QSO where necessary.    
The solid circles in Fig.~\ref{degenerate}
indicate the photometry of region $a$.   Clearly, the older model (red line)
is a better fit to the data.  
It is important to note, however, that if there is considerable reddening 
by dust along the lines of sight to these regions in
the host galaxy, the flux values of $U'$ will be depressed; 
therefore the photometric points can only help us to determine upper limits 
to the ages of the stellar populations.   For consistency,
the age errors quoted in this section are as defined in \S\ref{modeling},
and do not take into account additional constraints placed by the photometry.

Region $a$ corresponds to the west side of the arc-shaped 
structure \citep[the ``horseshoe'';][]{ham87},
$\sim4$\arcsec\ south of the QSO nucleus.  The {\it HST} F702W image
in Fig.~\ref{mrk231hst} shows that region $a$ is formed by 
multiple knots, much like those of regions $a$ and $c$ in
IRAS\,07598+6508.   However this structure, unlike those of IRAS\,07598+6508,
has a remarkably similar morphology from the near UV to the near IR as shown
in the top right panel of Fig.~\ref{mrk231mos}.  As discussed above, we
determine an age of 42 (+22, -17) Myr.   This age is younger 
than the 225 Myr estimated from $U\!-\!B$ colors of this region by 
\citet{sur00}.
 
While several authors have noted the blue color of the ``horseshoe''
\citep[\eg][]{kod79,hut87},
we find a region that is relatively 
much brighter at shorter wavelengths 16\arcsec\
south of the QSO, labeled $c$ in Fig.~\ref{mrk231slit}.   This region appears 
as a bright blue extended blob dominating the $U'\!-\!B$ color map in 
Fig.~\ref{mrk231mos}.  The spectrum has a very steep blue continuum, and 
very strong emission lines at the same redshift as the absorption lines,
likely from an H\,II region \citep{ham87}.
The top panel of Fig.~\ref{mrk231spec} shows the spectrum of region $c$ with 
the Balmer emission lines subtracted as described in \S\ref{modeling}.
We find an age for this region of 5 Myr.  As such a young starburst age 
is comparable to the expected duration of a typical individual starburst,
and we are likely observing a collection of several such starbursts, 
it is virtually impossible to distinguish between continuous star formation
and instantaneous bursts.   The age we find is, therefore, simply indicative
of ongoing star formation.
As in region $a$, we find degeneracy in the modeling of this spectrum,
with a second $\chi^{2}$ minimum at 100 Myr.   The near-UV continuum of
the older model, however, falls $2\sigma$ below the $U'$ photometric point.
Neither the deep optical ground-based images nor the {\it HST} image show 
evidence for stellar associations in region $c$.   This is somewhat 
surprising since, although the starburst population contributes only 1\% to 
the luminous mass along the line of sight, its flux amounts to 42\% of the 
total flux at 5000 \AA (rest wavelength), and even more at shorter wavelengths.

Regions $b$ and $e$ appear as single clumps in the images and have ages 
of 140 (+80, $-$70) Myr and 180 (+60, $-$80) Myr respectively.   The 
starburst populations dominate the spectra in these regions, contributing
$\sim75$\% of the total flux at rest frame 5000 \AA\ and up to 36\% of
the total luminous mass along the line of sight.
Regions $d$, $f$ (Fig.~\ref{mrk231spec}), $g$, and $h$ all show very similar 
spectra, as expected from the color maps (Fig.~\ref{mrk231mos}), with ages 
between 300 and 360 Myr.
The starbursts in these regions typically contribute only $\sim$50\% to the 
total flux and $\sim15$\% of the total luminous mass,
except for region $d$, for which these values are very similar to those of
regions $b$ and $e$.

We detect very weak continuum from the north tail (region $l$) at the 
$2.5 \sigma$ level.   The continuum is slightly blue, and there is a hint
of \ion{Mg}{1}$b$ at the redshift of the host galaxy.  This region corresponds
to the blue region in the $B\!-\!V$ color map, just before the condensation
at the end of 
the north tail.   It is possible that this region contains a knot of star
formation like those found in the tail of Mrk\,1014.

\citet{lip94} reports the presence of an extended \ion{Na}{1} {\it D} 
BAL in the west side of the host galaxy.   The only place where we see
evidence for this feature is in region $j$.   
The spectrum of $j$, uncorrected for QSO scattered light, shows a wide 
absorption feature ($\sim1900$ km s$^{-1}$
FWHM compared to $\sim1050$ km s$^{-1}$ for the QSO \ion{Na}{1} {\it D} BAL)
as well as narrow absorption and emission (see below) at the
redshift of the stellar component.   After correcting for QSO contamination,
the feature appears narrower, but still at $\sim-$6400 km s$^{-1}$ with
respect to the emission line.    The hypothetical BAL feature is,
however, much weaker than that of the QSO, and it could
be an artifact of an imperfect subtraction of the QSO light.

While the evidence for extended BAL is not strong in our data, \citet{ham87}
find an ``excess flux'' extending to the blue 
side of [\ion{N}{2}] $\lambda$6548, which they interpret as a broad 
blueshifted component of H$\alpha$ or H$\alpha$ +
[\ion{N}{2}] indicative of an outflow with velocities up to 1500 km s$^{-1}$.
We observe this excess most clearly in regions $h$ and $g$ extending to even
larger velocities ($\sim -2500$ km s$^{-1}$).
\citet{bor92} find some level of excess light in 
the blue wing of the H$\alpha$ line in every low-ionization BAL QSO in their
sample and suggest that this excess flux could be from the BAL material
itself.

The narrow \ion{Na}{1} {\it D} emission line shows a peculiar
behavior.   
We observe narrow \ion{Na}{1} {\it D} emission in regions $h$ and $j$ 
on the red side of the absorption line, forming what looks
like a P-cygni profile.    The only other emission lines evident in region 
$h$ are [\ion{N}{2}] $\lambda$6583, and weak [\ion{S}{2}] 
$\lambda\lambda$6717,6731, but these lines are slightly {\it blueshifted}
with respect to the absorption lines.   
In region $g$ the \ion{Na}{1} {\it D} emission line disappears, but the 
absorption line becomes very weak as well, so it is possible that the 
emission line is blueshifted into the absorption line; [\ion{N}{2}],
however, is at a higher redshift than in region $h$.   Thus, 
the very low ionization gas appears to be decoupled from the moderately low 
ionization gas in these regions.

\citet{hut87} find a ``green'' (\ie\ visible only
in their $G\!-\!B$ image), jet-like feature 
extending from near the nucleus
to the northeast.  They suggest it ``may be line emission in [\ion{O}{3}],
perhaps tracing ionizing radiation originating in the nucleus''.   Our
slit goes through the western side of where this feature would be 
\citep[see Fig.~2 in][]{hut87},
and region $j$ should sample the brightest part
of the ridge leading to the bright knot.   However, we find only weak 
[\ion{O}{3}] in the spectrum of this region, certainly much weaker than 
in other regions we sample.   Therefore it is unlikely that this ``green''
feature, if real, is due to ionized gas.

Images of Mrk\,231 show a greatly disturbed host galaxy both in large and
small scales (Fig.~\ref{mrk231mos}).   At large scales, two symmetric tails 
extend on the east
side of the nucleus for $\sim44$\arcsec\ or $\sim35$ kpc each, and there is
some very extended low surface brightness material east of the north
tail.   The nucleus is off-center on a bright ``plateau'' \citep{ham87}
20 kpc across that shows complex morphology, including linear 
jet-like structures (region $g$), and curved tail-like structures 
(region $d$).   Closer to the nucleus (Fig.~\ref{mrk231hst}) we find the 
``horseshoe'' 3 kpc to the south with numerous clumps of star formation, 
and arm-like features spiraling around the nucleus which connect to the 
outer tidal tails \citep{sur98}
This morphology is indicative 
of a merger between two disk galaxies of similar mass.

While we are unable to obtain stellar spectra within the central kpc
($<2\arcsec$) of Mrk\,231, \citet{smi95} have found from
UV spectropolarimetry of the nuclear region that high polarization in
the optical falls off quite rapidly shortward of $\sim3000$ \AA; this
effect is most easily interpreted as a dilution of the polarized component
by O and B stars from an ongoing starburst \citep{smi95}.
CO and radio observations also show evidence for
a centrally concentrated starburst \citep{dow98,tay99}.
It is possible that the knots along
the arm-like features around the nucleus have ages similar to those of 
region $a$, if not younger.

We find a relatively wide range  of starburst ages around the host galaxy.   
If, once again, we assume a projected velocity of 300 km s$^{-1}$, the tidal
tails have a dynamical age of 110 Myr.  Mrk\,231 is unusual in that there 
seems to be a significant post-starburst population somewhat
older (300--360 Myr) than the tails, which may indicate that wide spread
star formation was ignited prior to the stages of final merger.  Numerical 
simulations of mergers \citep{mih96}
predict such a scenario
when the merging galaxies lack a significant bulge to stabilize them against
an early dissipation.   In CS2000 we noted that the apparent correlation
between bulge mass and black-hole mass \citep{mag98}
suggests a possible correlation between QSO luminosity and 
the delay in star formation activity.   Thus, the early starburst activity 
in Mrk\,231 may indicate that the host galaxy did not have a substantial 
bulge, which
would in turn imply a less massive black-hole that may result into a less
luminous AGN.   Although we do in fact observe a relatively less luminous
AGN in Mrk\,231, these connections are highly speculative, and they are
based on relations \citep{mag98}
and models \citep{mih96}
that still have many uncertainties.

\section{Discussion}

Mrk\,1014, IRAS\,07598+6508, and Mrk\,231 show many similarities.
We already
knew that these three objects, while being optically or IR selected QSOs, 
were also part of the ULIG family.   In this study we have found 
additional properties which they share and which may be related to their 
intermediate position in the FIR diagram.

While Mrk\,1014 and Mrk\,231 have long been known to have tails, we
have shown that IRAS 07598+6508 also has at least one tidal tail.
We have found evidence that all three objects have undergone a strong
interaction and are now in the final stages of mergers.  The morphology
of the hosts (highly perturbed galaxies with tidal tails and destroyed disks)
as well as the extent of the starbursts
indicate that these are all major mergers between galaxies of comparable
mass rather than accretion events of low mass dwarf companions 
\citep{mih94,mih96}.

All three objects show spectra typical of E+A galaxies \citep{dre83},
that is, spectra characterized
by the simultaneous presence of strong Balmer absorption lines indicative 
of a young stellar population and features from an older population such 
as \ion{Mg}{1}{\it b}, and the absence of strong emission lines typical 
of star forming galaxies (we reiterate
that the emission lines seen in our spectra generally come from extended gas 
ionized by the QSO rather than from star forming regions, with 
a few exceptions). 
E+A galaxies are frequently linked to galaxy-galaxy
mergers and interactions \citep[\eg][]{zab96}.
In the case of objects with E+A spectra undergoing tidal interactions,
the young superposed population is clearly related to the interaction.
The three objects discussed in this paper, along with
3C\,48 (CS2000) and PG\,1700+518 (CS97), all have interaction-induced 
starbursts.

In addition to the similarities noted in the host galaxies of the
three objects, there are also some common characteristics in the 
spectra of the active nuclei themselves.  All three objects show
strong \ion{Fe}{2} emission, with two of these objects 
(IRAS\,07598+6508 and Mrk\,231) being ``extreme''
\ion{Fe}{2} emitters (\ie\ showing ratios of 
$I$(Fe\,II $\lambda4570$)/$I$(H$\beta$) $>$ 2; 
\citealt{lip94}).
The latter two objects, like PG\,1700+518
(CS97; SCC98), are also low-ionization BAL QSOs and have 
weak or absent narrow emission lines.

How do the ages of these starbursts relate to the merger/interaction
stage?   The dynamical ages for tidal features in these objects are
uncertain because of projection effects.
However, our rough estimates indicate
that in two cases, Mrk\,1014 and IRAS\,07598+6508, the peak of the
starburst occurred after the tidal tails were launched.   This requires
some mechanism to stabilize the gas contents of the galaxies against
bursting in star formation until the later stages of the merger; numerical
simulations indicate that a significant bulge in the host can provide this
mechanism \citep{mih96}.
Mrk\,231, on the other hand, shows starburst
ages which indicate that much of the star formation activity may have started 
before the last stages of the merger; hence, one or both of the merging 
galaxies may have lacked a significant bulge.

Whenever we have found a significant range of starburst ages in the host
galaxies of QSOs, we have also found evidence that the youngest major 
starburst regions are preferentially concentrated towards the center of
the galaxy.   In 3C\,48 (CS2000) we found the clearest example of this, 
with starburst ages becoming progressively younger and more dominating as we 
approached the galaxy/QSO nucleus.  Mrk\,231 shows older starbursts in the
``plateau'' extending $\sim12$ kpc around the nucleus, with some of the 
youngest
populations only 3 kpc from the nucleus, and possibly even younger populations
in the central kpc of the galaxy.   
Mrk\,1014 shows stronger relatively recent star formation activity along
the tail than any of the other objects.  
However, the starburst region we observe within 2 kpc of the
nucleus is far more massive, luminous, and larger \citep{sur98}
than the starburst regions along the tail, and has the youngest age found 
in the host.

The interaction histories of the five objects discussed so far
clearly favor a strong connection between interactions and vigorous 
bursts of star formation.   Since the gas flows towards the inner regions
can not only trigger star formation but also serve as fuel to the QSO,
one might expect the age of the QSO activity to be closely related to the age 
of the initial starbursts in the central regions of the galaxy.   However,
observations of central starbursts are in every case hampered by the 
presence of the QSO; even if this region could be observed, the spectrum 
would likely be dominated by continuing recent starburst activity and not 
by a starburst that was 
coincident with the onset of the QSO activity. Furthermore, both starburst
and QSO activity may be episodic.
All of this is to say that there is an unavoidable intrinsic 
uncertainty in using starburst ages
to place QSOs in an evolutionary sequence.  
We defer the detailed discussion of a possible age sequence to a
subsequent paper where we will present the results of the four remaining
objects in our sample (\ie\ IRAS\,00275$-$2859, IRAS\,04505$-$2958, I\,Zw\,1, 
and PG\,1543+489).

We emphasize, though, that many of our general conclusions from the objects
discussed here, as well as from our observations of 3C\,48 (CS2000), have
an interest that is quite independent of any attempt to use some sort of
starburst age as a proxy for a QSO age.  (1) The confirmation that these are
all starburst or post-starburst objects and that they all show obvious
tidal tails validates their close connection with other ULIGs, virtually all
of which are mergers or strongly interacting pairs.  While there has long
been strong {\it circumstantial} evidence that a large fraction of the QSO
population has resulted from triggering of the QSO activity by interactions
and mergers, we now have much more direct evidence for this mechanism for
at least one subclass of QSOs.  (2) Both the spatial distribution and
the time history of star formation in a QSO host galaxy give clues to
nature of the galaxies that have participated in the merger.  While more
sophisticated models of star formation during interactions will be necessary 
to exploit these data fully, we already have some hints in terms of the
enhanced star formation along the leading edge of the tails in 3C\,48 and
Mrk\,1014, and in the relative ages of the tail structures and the star-forming
regions contained within them.  (3) The youth of the stellar populations in
these objects reinforces previous suggestions connecting strong \ion{Fe}{2}
emission and low-ionization BAL features with the relatively recent triggering
of QSO activity.  We will discuss these connections in detail in the
paper presenting the observations of the remaining four objects in our sample.

\acknowledgments

We thank Gerbs Bauer, Scott Dahm, and Susan Ridgway for assisting in some of 
the observations, and Bill Vacca for helpful discussions about IMFs.
We also thank the referee, Dean Hines, for his very prompt review of
the paper and his suggestions, which helped us improve both its content
and its presentation.
This paper was partly written while both authors were visitors at the
Research School of Astronomy and Astrophysics of the Australian National
University, and we thank both the Director, Jeremy Mould, and the staff 
there for their hospitality.
This research has made use of the NASA/IPAC Extragalactic Database (NED) 
which is operated by the Jet Propulsion Laboratory, California Institute of 
Technology, under contract with the National Aeronautics and Space 
Administration.   This research was partially supported by NSF under grant 
AST95-29078.

\clearpage

\figcaption{Morphology and colors in the host galaxy of Mrk\,1014.  The main
panel on the left shows a $R'$-band image, heavily smoothed to show the
low-surface-brightness features; the upper-left inset shows the well known
local interacting pair M51/NGC\,5195, scaled to the distance of Mrk\,1014,
to indicate the size of the Mrk\,1014 host galaxy.  A lower-contrast version 
of the same Mrk\,1014 image
with less smoothing and with the PSF subtracted is shown at the middle top.
$K'$ and $B'$-band images are shown at the top right and bottom middle,
respectively, and the $B'\!-\!R'$ color image is shown at the bottom right.
The latter is coded so that regions with most negative $B'\!-\!R'$ show as
blue, while relatively redder regions show as yellow and red.  The region
near the nucleus is dominated by the QSO and does not indicate the color
of the host galaxy.\label{mrk1014mos}}

\figcaption{{\it HST} WFC2 image of Mrk\,1014 obtained with the F675W filter.
For this and the following {\it HST} images, the image has been rotated so 
that north is at the top and east to the left.\label{mrk1014hst}}

\figcaption{Slit positions and region identifications for 
Mrk\,1014.\label{mrk1014slit}}

\clearpage
\figcaption{Morphology in the host galaxy of IRAS\,07598+6508.  The main
panel on the left and the top two panels on the right show the $R'$ image
at a range of contrasts, while the lower panels on the right show the
$U'$ and [O\,III] images.  All images have had the PSF removed, using the
bright star to the north-west as the PSF template.
\label{ir0759mos}}

\figcaption{{\it HST} PC1 image of IRAS\,07598+6508 obtained with the F702W 
filter.\label{ir0759hst}}

\figcaption{Slit positions and region identifications for IRAS\,07598+6508.
\label{ir0759slit}}

\clearpage
\figcaption{Morphology and colors in the host galaxy of Mrk\,231. 
The main panel
on the left and the top two panels on the right show the $B$-band image.  The
insert in the top-right panel shows an $H$-band image of the inner regions
(note that the images in this panel are magnified by a factor of two compared
with those in the other panels);
the similarity in the structure seen in the arc-like region south of the
nucleus indicates that the SEDs of the emission regions within this region
are all similar and that reddening is not a severe problem.  
The two lower-right
panels show \bv\ and \ub\ images, respectively; more negative values are
indicated as blue.  Note the relative colors
of the arc and the patch about 16\arcsec\ south of the nucleus (essentially
regions $a$ and $c$ in Fig.~4), indicating the much younger 
population in the latter.\label{mrk231mos}}

\figcaption{{\it HST} PC1 image of the inner region of Mrk\,231 obtained
with the F439W filter.\label{mrk231hst}}

\figcaption{Slit positions and region identifications for Mrk\,231.
\label{mrk231slit}}

\clearpage

\begin{center}
\begin{deluxetable}{lclccc}
\tablecaption{Journal of Spectroscopic Observations \label{journal}}
\tablehead{\colhead{} & \colhead{PA} & \colhead{Offset}
& \colhead{Dispersion} & \colhead{Total Int.} & \colhead{} \\
\colhead{Object} & \colhead{(deg)} & \colhead{(arcsec)}
& \colhead{(\AA\ pixel$^{-1}$)} & \colhead{Time (s)}& \colhead{UT Date} }
\startdata
Mrk\,1014 &  103.0   & \phn\phn4.0 N  & 2.44 & 3600 & 97 Sep 13  \\
Mrk\,1014 & \phn87.1 & \phn\phn6.9 N  & 2.44 & 3600 & 98 Sep 01  \\
Mrk\,1014 &  148.6   & \phn\phn2.2 W  & 2.44 & 2400 & 98 Sep 01  \\
IRAS\,07598+6508 & 243.6 & \phn\phn6.9 S & 1.28 & 3600 & 96 Oct 13 \\
IRAS\,07598+6508 & 190.0 & \phn\phn0.0   & 1.28 & 3600 & 96 Oct 14 \\
Mrk\,231 & \phn\phn6.0  & \phn\phn0.0 & 2.44 & 1800 & 97 Jun 12  \\
Mrk\,231 & \phn83.0  & \phn16.0 S     & 2.44 & 3600 & 98 Mar 21  \\
Mrk\,231 & \phn30.0 &  \phn\phn7.0 W  & 2.44 & 2400 & 98 Mar 21  \\
\enddata
\end{deluxetable}
\end{center}

\begin{center}
\begin{deluxetable}{llcccc}
\tablecaption{Journal of Imaging Observations \label{imaging}}
\tablehead{\colhead{Object} & \colhead{Instrument} & \colhead{Filter\tablenotemark{a}}
& \colhead{Exposure (s)}& \colhead{Scale (\arcsec pix$^{-1}$)} 
& \colhead{UT Date} }
\startdata
Mrk\,1014 & Orbit CCD   & $U'$        &\phn8$\times$1200     & 0.138 & 97 Nov 01  \\
Mrk\,1014 & Orbit CCD   & $B'$        &\phn5$\times$300\phn  & 0.138 & 97 Oct 31  \\
Mrk\,1014 & Tek1024 CCD & $R'$        &\phn5$\times$300\phn  & 0.222 & 91 Nov 13  \\
Mrk\,1014 & QUIRC       & $H$         & 29$\times$300\phn    & 0.061 & 99 Oct 29  \\
Mrk\,1014 & NICMOS-3    & $K'$        & 39$\times$70\phn\phn & 0.374 & 90 Oct 28  \\
IRAS\,07598+6508&Loral CCD&$U'$       &\phn8$\times$1200     & 0.138 & 97 Mar 09  \\
IRAS\,07598+6508&Tek1024 CCD&$R'$     & 15$\times$300\phn    & 0.222 & 92 Mar 01  \\
IRAS\,07598+6508&Tek1024 CCD&[O\,III] &\phn7$\times$1560     & 0.222 & 92 Feb 27 \\
IRAS\,07598+6508& QUIRC & $H$         & 22$\times$120\phn    & 0.061 & 99 Oct 30  \\
Mrk\,231  & Loral CCD   & $U'$        &\phn5$\times$1200     & 0.138 & 97 Mar 08  \\
Mrk\,231  & Loral CCD   & $B$         &18$\times$200\phn     & 0.138 & 97 Mar 09  \\
Mrk\,231  & Loral CCD   & $V$         &\phn6$\times$300\phn  & 0.138 & 97 Mar 08  \\
Mrk\,231  & QUIRC       & $H$         &69$\times$20\phn\phn  & 0.061 & 99 Apr 04 \\
\enddata
\tablenotetext{a}{The $U'$ filter was centered at 3410 \AA\ with a bandpass 
of 320 \AA .  The $B'$ filter used for Mrk\,1014 had a smaller bandpass
(593 \AA ) than the standard $B$ filter to avoid strong emission lines. 
The $R'$ filter used for Mrk\,1014 
was centered at 6815 \AA\ with a bandpass of 1226 \AA, and that used for 
IRAS\,07598+6508 was centered at 6500 \AA\ with a bandpass of 880 \AA; 
both avoid strong emission lines.  The [O\,III] filter was centered at 
5777 \AA\ with a bandpass of 100~\AA.}
\end{deluxetable}
\end{center}

\end{document}